\documentclass{mn2e}
\usepackage{times}
\usepackage{amsmath}
\usepackage{amssymb,amsfonts}
\usepackage{graphicx}
\usepackage{epstopdf}
\usepackage{verbatim}

\def\gapprox{\lower.4ex\hbox{$\;\buildrel >\over{\scriptstyle\sim}\;$}}
\def\lapprox{\lower.4ex\hbox{$\;\buildrel <\over{\scriptstyle\sim}\;$}}
\def\be{\begin{equation}}
\def\be{\begin{equation}}
\def\ee{\end{equation}}
\def\bea{\begin{eqnarray}}
\def\eea{\end{eqnarray}}
\catcode`\@=11 \font\tenmib=cmmib10 \font\tensyb=cmbsy10
\font\tenbi=cmmib10
\newfam\bifam 
\textfont\bifam=\tenbi  %for bold italics; textstyle only

\def\unboldmath{\everymath{}\everydisplay{}
          \textfont\@ne\teni
          \textfont\tw@\tensy
          }
\def\boldmath{$\!\!$\relax\everymath{\mit}\everydisplay{\mit}
        \textfont\@ne\tenmib
        \textfont\tw@\tensyb
        \relax}%
\catcode`\@=12

\begin{document}
\tolerance=10000

\def\lesssim{\mathrel{\hbox{\rlap{\hbox{\lower4pt\hbox{$\sim$}}}\hbox{$<$}}}}
\def\gtrless{\mathrel{\hbox{\rlap{\hbox{\lower3pt\hbox{$<$}}}\hbox{$>$}}}}

\def\cenx4{{Cen~X$-$4}}
\def\aql{{Aql~X$-$1}}
\def\1e{{1E 1207.4-5209}}
\def\exo{{EXO 0748-676}}
\def\saxj{{SAX J1808.4$-$3658}}
\newcommand{\ud}[2]{\mbox{$^{+ #1}_{- #2}$}}

%\draft
\def\be{\begin{equation}}
\def\ee{\end{equation}}
\def\bea{\begin{eqnarray}}
\def\eea{\end{eqnarray}}
\def\c{\cite}

\def\et{ {\it et al.}}
\def\lan{ \langle}
\def\ran{ \rangle}
\def\ov{ \over}
\def\ep{ \epsilon}

\def\et{ {\it et al.}}
\def\la{ \langle}
\def\ra{ \rangle}
\def\ov{ \over}
\def\ep{ \epsilon}

\def\mdot{\ifmmode \dot M \else $\dot M$\fi}    % accretion rate
\def\mxd{\ifmmode \dot {M}_{x} \else $\dot {M}_{x}$\fi}
\def\med{\ifmmode \dot {M}_{Edd} \else $\dot {M}_{Edd}$\fi}
\def\bff{\ifmmode B_{f} \else $B_{f}$\fi}

\def\apj{\ifmmode ApJ\else ApJ\fi}    % lower
\def\apjl{\ifmmode  ApJ\else ApJ\fi}    %
\def\aap{\ifmmode A\&A\else A\&A\fi}    %
\def\mnras{\ifmmode MNRAS\else MNRAS\fi}    %
\def\nat{\ifmmode Nature\else Nature\fi}
\def\prl{\ifmmode Phys. Rev. Lett. \else Phys. Rev. Lett.\fi}
\def\prd{\ifmmode Phys. Rev. D. \else Phys. Rev. D.\fi}

\def\ms{\ifmmode {\rm M_{\odot}} \else ${\rm M_{\odot}}$\fi}    % lower
\def\na{\ifmmode \nu_{A} \else $\nu_{A}$\fi}    % Alfven frequency
\def\nk{\ifmmode \nu_{K} \else $\nu_{K}$\fi}    % Keplerian frequency
\def\ns{\ifmmode \nu_{{\rm s}} \else $\nu_{{\rm s}}$\fi}
\def\no{\ifmmode \nu_{1} \else $\nu_{1}$\fi}    % lower
\def\nt{\ifmmode \nu_{2} \else $\nu_{2}$\fi}    % upper
\def\ntk{\ifmmode \nu_{2k} \else $\nu_{2k}$\fi}    % upper
\def\dnmax{\ifmmode \Delta \nu_{max} \else $\Delta \nu_{2max}$\fi}
\def\ntmax{\ifmmode \nu_{2max} \else $\nu_{2max}$\fi}    % upper
\def\nomax{\ifmmode \nu_{1max} \else $\nu_{1max}$\fi}    % upper
\def\nn{\ifmmode \nu_{\rm NBO} \else $\nu_{\rm NBO}$\fi}    % HBO
\def\nh{\ifmmode \nu_{\rm HBO} \else $\nu_{\rm HBO}$\fi}    % HBO
\def\nqpo{\ifmmode \nu_{QPO} \else $\nu_{QPO}$\fi}    % HBO
\def\nz{\ifmmode \nu_{o} \else $\nu_{o}$\fi}    % HBO
\def\nht{\ifmmode \nu_{H2} \else $\nu_{H2}$\fi}    % HBO
\def\ns{\ifmmode \nu_{s} \else $\nu_{s}$\fi}    % stellar
\def\nb{\ifmmode \nu_{{\rm burst}} \else $\nu_{{\rm burst}}$\fi}
\def\nkm{\ifmmode \nu_{km} \else $\nu_{km}$\fi}    % stellar
\def\ka{\ifmmode \kappa \else \kappa\fi}    % stellar
\def\dn{\ifmmode \Delta\nu \else \Delta\nu\fi}    % stellar

\def\vk{\ifmmode v_{k} \else $v_{k}$\fi}    % Keplerian velocity
\def\va{\ifmmode v_{A} \else $v_{A}$\fi}    %
\def\vf{\ifmmode v_{ff} \else $v_{ff}$\fi}    % free fall velocity

\def\rs{\ifmmode {R_{s}} \else $R_{s}$\fi}    % stellar
\def\ra{\ifmmode R_{A} \else $R_{A}$\fi}    % Alfven radius
\def\rso{\ifmmode R_{S1} \else $R_{S1}$\fi}    % sonic point radius
\def\rst{\ifmmode R_{S2} \else $R_{S2}$\fi}    % sonic point radius
\def\rmm{\ifmmode R_{M} \else $R_{M}$\fi}    % stellar
\def\rco{\ifmmode R_{co} \else $R_{co}$\fi}    % stellar
\def\ris{\ifmmode {R}_{{\rm ISCO}} \else $ {\rm R}_{{\rm ISCO}} $\fi}
\def\rsix{\ifmmode {R_{6}} \else $R_{6}$\fi}
\def\rinfty{\ifmmode {R_{\infty}} \else $R_{\infty}$\fi}
\def\rinfsix{\ifmmode {R_{\infty6}} \else $R_{\infty6}$\fi}

\def\rxj{\ifmmode {RX J1856.5-3754} \else RX J1856.5-3754\fi}

\title{Measuring Neutron Star Mass and Radius with Three Mass-Radius
Relations}
%\titlerunning{Measuring Neutron Star Mass and Radius}

\author[C.M. Zhang et al.]{C. M. Zhang$^{1}$, H.X. Yin$^{1}$,
 Y.  Kojima $^{2}$,
 H. K. Chang $^{3}$,
 R. X. Xu $^{4}$,
 X. D. Li $^{5}$,
 B. Zhang $^{6}$,
 B. Kiziltan $^{7}$\\
\\
1. National Astronomical Observatories,
 Chinese Academy of Sciences,
Beijing 100012, China\\
2.  Department of Physics, Hiroshima University,
Higashi-Hiroshima 739-8526, Japan\\
3. Department of Physics and Institute of Astronomy, National
Tsing Hua University, Hsinchu 30013, Taiwan, China\\
4. Department of Astronomy, Peking University, Beijing 100871,
China\\
5.  Department of Astronomy, Nanjing University, Nanjing 210093,
China\\
6.  Department of Physics, University of Nevada, NV 89154-4002, USA
\\
7. Department of Astronomy \& Astrophysics, University of
California, Santa Cruz, USA}

\date{\today}

% {The 2nd reply to the referee is enclosed}
 \maketitle

 \begin{abstract}
We propose to determine the mass and the radius of a neutron star
(NS) using three measurable mass-radius relationships, namely the
``apparent'' radius inferred from neutron star thermal emission, the
gravitational redshift inferred from the absorption lines, as well
as the averaged stellar mass density inferred from the orbital
Keplerian frequency derived from the kilohertz quasi periodic
oscillation (kHz QPO) data. We apply the method to constrain the NS
mass and the radius of the X-ray sources, 1E 1207.4-5209, Aql X-1
and EXO 0748-676.
\end{abstract}


\section{Introduction}

Measuring the mass ($M$) and the radius $R$ of a neutron star (NS)
is very important in both nuclear physics and gravitational physics
since it allows us to  constrain the NS matter compositions (e.g.
neutrons or quarks) at super-density (see, e.g. Lattimer \& Prakash
2004; Miller  2002; Li et al. 1999; Cheng et al. 1998; Haensel et
al. 1986; Alcock et al. 1986), and to trace the particle motion
behavior in super-strong gravitational field (see, e.g. van der Klis
2000, 2006).

NS masses  have been measured  in the radio pulsar binary systems
with high precision (see, e.g. Kaspi et al. 1994; Bailes et al.
2003). For example, the masses of the two NSs of the Hulse-Taylor
pulsar system, PSR 1913+16, have been measured as
%with the   masses
M=1.41$\ms$ and M=1.38 $\ms$, respectively (see, e.g., Manchester
\& Taylor 1977); and those of the recently discovered double
pulsar system, PSR J0737-3039, are
% with the masses
M=1.34$\ms$ and M=1.25 $\ms$, respectively (see, e.g., Lyne et al.
2004).
%  Furthermore,
The  mass constraints  of 26 neutron stars in the binary radio pulsar
systems have been derived  by Thorsett \& Chakrabarty (1999),
who present  a remarkably narrow underlying gaussian mass
distribution at  M=1.35$\pm0.04\ms$.
%  Moreover,
In the low mass X-ray binaries (LMXBs)
%  the  NS masses have also been measured to be
a lower mass limit $M > 0.97\pm0.24\ms$ was derived for the NS in 2A
1822-371
%as a lower limit
(Jonker et al 2003), and the NS mass in Cyg X-2 was derived as
$M=1.78\pm0.23 \ms$
%in   Cyg X-2
(Orosz \& Kuulkers 1999).
%  respectively.

In contrast, there is no method so far to
%However,   the
directly measure the radius of a NS.
%has not yet
%existed. As the conventional estimations, the NS radii  are
A conventional value of NS radius is about 15 km, which is derived by
% approximately assumed to be about 15 (km) to concord
requiring that the averaged dipolar magnetic field
 strength (i.e. $\sim 10^{12}$ G) inferred from radio pulsar spindown
 (see, e.g. Manchester \& Taylor 1977) is
consistent with the one implied by the cyclotron line emission in
some X-ray binaries
% with those
% of X-ray neutron stars implied by the
% cyclotron  resonance lines
(see, e.g. Makishima et al. 1999; Truemper et al. 1978).
% Coburn et al. 2002, and references therein).
% Therefore,
The uncertainties in determining NS
 radius make it difficult to accurately measure the NS
equation-of-state (EOS).  The detail of matter composition inside a
NS has been an open issue.

With the launches of X-ray space observatories, especially  the
recent XMM-Newton, Chandra, and  RXTE (see, e.g. Jansen et al. 2001;
van der Klis 2000, 2006), some $M-R$ relationships have been
appropriately measured or estimated. These include (1) the
``apparent radius" $\rinfty$ estimated from the thermal emission of
isolated NSs (e.g. Geminga, Caraveo et al.  2004; 1E 1207.4-5209,
Bignami et al. 2004; RX J1856.5-3754, Truemper et al. 2004; Burwitz
et al. 2003; Burwitz et al. 2001) or NSs in binary systems (e.g. Aql
X-1, Rudledge et al. 2001); (2) the gravitational redshift implied
from the absorbtion lines in EXO 0748-676 and in 4U 1700+24  (e.g.,
Cottam et al. 2002; Tiengo et al 2005)
%that infers the ratio of NS mass to radius (see, e.g., Miller 2002),
%as well as
and (3) the averaged NS mass density inferred
from the orbital Keplerian frequency used to interpret
% as an  interpretation of
the kHz QPOs discovered by  RXTE (see, e.g., Miller et al. 1998; van
der Klis 2000, 2006; Zhang 2004).

Generally speaking,
%the above ``measured'' M-R relations have provided us with the
%constraints on the NS masses and radii,
% and in principle we   should definitively give the
if two of the three above $M-R$ relations are well ``measured'', one
can solve both $M$ and $R$ independently.
% independent values of M and R if  two M-R relations are
% conveniently  known.
This is the main topic of this paper. We will discuss the methods in
detail and apply them to several sources. {\bf Once NS $M$ and $R$
known, we can evaluate the matter compositions of star by comparing
with the representative EOSs, as shown in the three figures we
chosen the EOSs of strange matter (CS1 and CS2), normal neutron
matter (CN1 and CN2) and pion condensate in the star core (CPC)
(see, e.g. Lattimer \& Prakash 2001; Cook et al. 1994). }
%Hence,  the purpose of this letter is to exploit
%  these ``measured"  M-R relations to determine or constrain
%   the independent values of  M and R, the derivation of
% which, together with some applications,  is  described in section 2, and
% the summary is given in the last section. As a conventional
%usage,
%Following the conventions, both the gravitational constant G and the
%The speed of light $c$ is taken as unity throughout the Letter.

\section{Measuring NS mass and radius with $M-R$ relations}

%At first,

  (1) The  ``apparent radius" $\rinfty$ inferred from the
thermal emission data is only an upper limit of the NS radius. It is
related to the ``true"  radius $R$ through (see, e.g., Haensel 2001;
Lattimer \& Prakash 2001; Thorne 1977), \be R_{\infty} =
R/\sqrt{1-\rs/R} \;,\label{rinfty} \ee where $\rs=2GM$ is the
Schwarzschild radius. From  Eq.(\ref{rinfty}), the maximum value of
the mass can be easily obtained as  ${M=1.3(\rinfty/10^{6}cm)\ms}$
when $R=0.58 \rinfty$.  {\bf We stress that the apparent radius is
not a directly observable quantity although derivable (in principle)
from two observable quantities, namely the "apparent blackbody
luminosity" (the blackbody luminosity
 seen by a distant observer) and the "apparent blackbody temperature".
 Both  quantities are derived by fitting X-ray data under the
 assumption of perfect blackbody, which is, in general, not true
(see, Haensel 2001). In particular the inclusion of an atmosphere
 above the NS crust can vary the estimate of the NS apparent radius
 by more than a factor 2 (see Haensel 2001, section 3.1), thus the
  `measured' apparent radius is still model dependent.
}

%In practice, however, the precision of calculating
%the ``apparent radius"  is not as high as that of
%measuring the NS mass,
%since the ``measurement"  is affected by many factors, such as
%uncertainties of the distance of the source, deviation of the simple
%blackbody spectrum of the thermal emission, etc. (e.g. Rutledge et
%al. 2001; Haensel 2001).
%the estimations of the apparent
% luminosity,  the apparent temperature and  the distance of the
% source to the observer, etc (see, e.g., Rutledge et al. 2001;
% Haensel 2001).

%Secondly,
(2) The gravitational redshift near the neutron star surface
%caused by the strong gravitational field at the star surface
provides another M-R relation, giving the ratio between the NS mass
and the NS radius (e.g. Miller 2002). The approach becomes relevant
with the recent discovery of
%and recently  the approaches have  existed
%to measure   it  in
the absorption features  in  EXO 0748-676 (Cottam et al. 2002).
%  Mathematically, the gravitational  redshift z   is   described by
%the relative shift of the spectrum  wavelength  $\lambda$  as
According to the definition of the gravitation redshift
$z={\Delta \lambda/\lambda} \simeq{GM/R}$ (for a weak
gravitational field, where $\lambda$ is the wavelength of the
emission),
%then,
for a  non-rotating spherical gravitational source in  the
Schwarzschild spacetime, we get
%usually  write
% the gravitational redshift formula in the following
(see, e.g.,  Sanwal et al. 2002),
 \be
 1+z = 1/\sqrt{1-\rs/R}\;,
  %(1-{2GM\over R})^{-1/2}\;, %=(1-{0.3m\over R_{6}})^{-1/2}\,,
\label{redshift}
\ee
or, in  terms of  $M/R$,
 \be
 {m/R_{6}} = f(z) = {z(1+z/2)\ov 0.15(1+z)^{2}}\,,
 \label{redshift2}
\ee
where ${\rm R_{6}=R/10^{6}cm}$, and $m={\rm M/\ms}$ is the mass in
unit of solar mass. In the case of weak gravitational fields
($z \ll 1$), $f(z) \simeq {z/0.15}$ is approximately obtained.

(3) A third $M-R$ relation can be obtained by analyzing the kilohertz
quasi periodic oscillation (kHz QPO) data.
% or the millisecond
%variability of the LMXB systems.
%the X-ray spectra of LMXBs  QPO data.
This field has been greatly benefited from the RXTE mission,
%Thirdly,  with
% the launch of   RXTE,  it  has  provided  us with the opportunity
% to acquire   an estimation of the averaged mass
%density  ${\rm M/R^{3} }$   of neutron star. In order  to probe
%the orbital motion around the
% accreting NS  in the binary system,  RXTE has been pushed
% into space and
which discovered
the twin (upper and lower) kHz QPOs
%kilohertz quasi periodic oscillations (kHz QPOs)
or millisecond variability in the X-ray data of about 20 NS LMXB
systems.
%have been which discovered  in  about twenty NS X-ray  sources
%(see, e.g. van der Klis 2000, 2004).
%As the conventional mechanism arguments,
The upper kHz QPO frequency is generally interpreted as the
Keplerian frequency  $\nk$ of orbital materials at some preferred
radius $r$ near the NS surface (van der Klis 2000, 2006; Morsink
2000; Stella \& Vietri 1999; Zhang 2004), i.e., \be \nk =
\sqrt{{GM\over 4\pi^{2}r^{3}}} = 1850~ {\rm (Hz)}
{}A{}X^{3/2}\,,\,\,\, X=R/r \, ,\label{nt} \ee with the parameter $A
= (m/R_{6}^{3})^{1/2}$ or $R_{6} = 1.27{}m^{1/3}(A/0.7)^{-2/3}$,
where $R_6$ is the NS radius in unit of 10 km. The physical meaning
of the parameter A is such that
%,  it is explicitly  that
the quantity $A^{2}$ represents  the ``measurement"  of the averaged
NS mass density, namely,
\be
{M/R^{3} \simeq 10^{15} ({\rm g/cm^{3}}) (A/0.7)^{2}}\;. \label{mr}
\ee

Meanwhile, by interpreting the twin kHz QPOs as the
Keplerian motion and the Alfv\'en wave oscillation mechanism
(see, e.g., Zhang 2004), the averaged mass density parameter A can
be inferred from the simultaneously detected twin kHz QPO
frequencies. For typical kHz QPO sources, $A\simeq0.7$ is
approximately obtained.
%for the typical kHz QPO sources.
If, instead, only one single kHz QPO
frequency is detected, one can not derive the exact value of $A$ from
the model,
%we do not know the exact value of parameter
%A  from the model  but we can
but can estimate the lower limit of $A$ and the upper limit of $M$
%stellar mass in
according to the following arguments, as proposed by Miller et al.
(1998). In order to explain the observed X-ray flux saturation
%phenomena  in
at the twin kHz QPO frequencies
%at a certain X-ray flux level (see, e.g.,
(e.g. van der Klis 2000, 2006; Zhang et al. 1998), it is usually
believed that the maximum Keplerian frequency $\nk$ occurs either at
the star surface ($r=R$: $X=1$) or at the innermost stable circular
orbit (ISCO) with the radius ${\ris=3\rs=6GM}$. The two inequalities
are therefore derived from Eq.(\ref{nt}) of Miller et al (1998),
i.e. \be A \ge \nk/1850 ({\rm Hz})\;,\label{a} \ee and \be m \leq
{\rm 2200 (Hz)/\nk}\;. \label{mass2} \ee In addition, Burderi \&
King (1998) have derived a $M-R$ relation for the millisecond
accretion-powered X-ray pulsar SAX J 1808.4-3658 with the measured
spin frequency 401 Hz through comparing the stellar radius with its
co-rotation radius of the magnetosphere-disk. They inferred a value
of  $A>0.62$.

To summarize,
%roughly speaking, as a scope of our knowledge at moment,
hitherto there are roughly
%have existed
three kinds of well-known
$M-R$ relationships that could be measured or estimated from the data,
as described above.
%measurements or estimations as described above.
Knowing two of the three in a system would in principle lead to
the determination of both $M$ and $R$. In reality, errors are
large for these relationships. Knowing all three relations for a
source would greatly diminish the errors. As an illustration,
below we will use the method to constrain $M$ and $R$ for several
sources.

\subsection{
%Evaluation of  NS  mass and  radius by
Constraining $M$ and $R$ with $\rinfty$ and z:
% application
the case of 1E 1207.4-5209}

If the ``apparent radius" $\rinfty$  and the gravitational
redshift $z$ are known, both $M$ and $R$ can be calculated by the
following two formulae, derived from Eqs.(\ref{rinfty}) and
(\ref{redshift}),
 \be \rsix=\rinfsix/(1+z)\;, \label{riz}\ee
\be m= f(z) \rinfsix/(1+z)\;, \label{miz}\ee
%m=3.3\rinfsix[1-(1+z)^{-2}]/(1+z)\;. \label{miz}\ee
with $\rinfsix=\rinfty/10^{6}$cm.

In the spectral analysis of 1E 1207.4-5209, Bignami et al. (2004)
found that the best fitting continuum model includes two blackbody
components, i.e. a cooler component with
 a temperature $KT=0.163 \pm 0.003$ keV and an emitting radius of
$R_e = 4.6 \pm 0.1$ km, and a hotter component with
 $KT=0.319 \pm 0.002$ keV and $R_e = 0.83 \pm 0.03$ km.
If we take the cooler component as the emission from the whole star,
 then the cooler emitting radius 4.6$\pm$0.1 km can be regarded as
 a measured lower limit of the ``apparent radius" $\rinfty$.
 {\bf In addition, an emitting radius R=$4.6\pm0.1 (d/2kpc)$ km
in  1E 1207.4-5209 is obtained by the best fitting continuum model
of  the blackbody function and by the distance of $2kpc$ (Bignami et
al. 2004), however as claimed by Haensel (2001) that an emitting
radius may be doubled if the atmosphere model of the photon spectrum
is considered,
 i.e. the emitting radius may be as high as $\sim$9.4 km.  }

%Nevertheless,
On the other hand, two absorption features in the spectrum
of 1E 1207.4-5209 were detected with the {\em Chandra} ACIS
%detector aboard the Chandra X-Ray Observatory
(Sanwal et al. 2002). Interpreting these lines as Helium atomic
lines, a gravitational redshift is inferred as $z \sim (0.12-0.23)$
(Sanwal et al. 2002).  This conclusion is, however, inconclusive,
since the lines may be due to the cyclotron origin (e.g. De Luca et
al. 2004). Here we take this gravitational redshift measurement as
an unconfirmed, tentative case.

\begin{figure}
%\includegraphics[height=0.40\textwidth,angle=0]{f1.eps}
%\epsscale{0.7}
%\vskip 6.5cm \hskip   -0.40cm \epsfxsize=7.cm
%\special{psfile=f1.eps hscale=90 vscale=90 angle= 0} \hskip -1.5cm
\includegraphics[width=9cm]{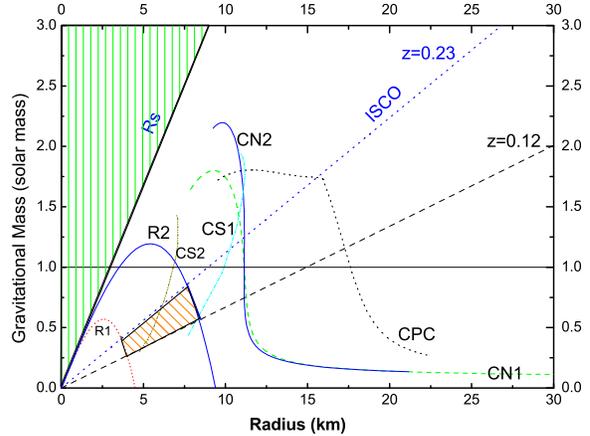}
 \caption[fig1]
 {\bf The $M-R$
diagram. Five representative EOSs are shown (see also, e.g. Miller
2002): stars containing strange quark matter (CS1 and CS2), stars
made of normal neutron matter (CN1 and CN2), and stars with pion
condensate cores (CPC) (Lattimer \& Prakash 2001; Cook et al. 1994).
The straight lines are constant $z$ lines: the solid ($\rs$) and the
dotted lines (ISCO) represent that the stellar radius is one and
three times of the Schwarzschild radius, i.e.  $R=2GM$ (below the
line $\rs$ the NS $M-R$ regimes are possible, see e.g. Haensel et
al. 1986) and $R=6GM$, respectively. The constant redshift lines
$z=0.12$ and $z=0.23$ are also plotted. The concave-down parabolas
are constant $\rinfty$ lines: $\rinfty=4.5, 9.4, 18$ km labeled by
R1, R2 and R3 correspond   the lower limit of the apparent radius of
1E1207.4-5290 (4.5=4.6-0.1) inferred
 by the blackbody assumption (Bignami et al. 2004),
 (9.4=2*(4.6+0.1) estimated by
taking into account the factor 2 uncertainty because of the
inclusion of the NS atmosphere (Haensel 2001)), and by considering
the possible longer distance from the source to the earth $d\sim3.9
kpc$ (see, e.g. Pavlov et al 2004), respectively. The shadowed area
 stands for the possible $M-R$ range of 1E1207.4-5209, which covers
the strange matter and the normal neutron EOSs. } \label{iz}
\end{figure}

{\bf In Figure \ref{iz}, we present the $M-R$ diagram of NSs with
various EOSs and present the above two $M-R$ constraints in the same
plot. For the possible  parameter ranges of $\rinfty$ and
$z=0.12-0.23$, i.e. $\rinfty=4.5-4.7 $ (km) as estimated from the
blackbody assumption (Bignami et al 2004), $\rinfty=9.4$ (km) while
the atmosphere model of photon spectrum is taken into account
(Haensel 2001), and $\rinfty=20$ (km) assumed under the condition
that the uncertainty in the distance of the source 1E1207.4-5209 to
the earth is considered (Truemper 2005),
  we constrain the parameter space within the area enclosed by the
lines labeled R1, R2, R3,  and those labeled z=0.12, 0.23, where the
implied $M-R$ ranges are given.
 In Figure \ref{iz}, we also  plot the lines of
 ${R=\rs=2GM}$ (the solid straight line labeled as $\rs$,
  below which the $M-R$ values will be permitted for NSs) and
${R=\ris=6GM}$ (the middle straight dotted line labeled as ISCO).
The latter corresponds $z=0.23$ (this value is fortuitous
coincidence with the observationally inferred redshift upper limit
of 1E1207.4-5209, see Sanwal et al. 2002) and divides the $M-R$
diagram into
  the $R<\ris$ region (above the line) and the $R>\ris$ region (below
the line).
  From Figure \ref{iz} the  inferred $M-R$ value regime for 1E1207.4-5209
  favors the existence of the exotic matter inside NS.
 However, it is pointed out by  Truemper (2005) that the distance
 $d$
from  1E 1207.4-5209 to the earth is  not so certain and  $d=1.3\sim
3.9 kpc$ has been given (see, e.g. Pavlov et al. 2004). If  we
consider the atmosphere model of photon spectrum and the distance
$d=3.9kpc$, then the derived ``apparent radius"  can be as large as
$18km$ (4.6*2*3.9/2=18), which will weaken the conclusion of 1E
1207.4-5209 including a strange star because the inferred $M-R$
 range in  Figure \ref{iz} is enlarged to cover the EOSs of
normal neutrons.

 If one assumes the validity of the above measured $\rinfty$ (4.5 -
4.7 km) and $z$ (0.12 - 0.23), then one gets $M=0.34\pm0.09\ms$ and
$R=4.2\pm 0.1$ km for 1E 1207.4-5209 according to Eqs.(\ref{riz})
and (\ref{miz}).  In this case, one can conclude the NS EOS of 1E
1207.4-5209 infers the composition of strange quark matters (e.g. Xu
2005).  Apparently, this conclusion is model dependent and based on
the blackbody assumption and the distance of $2kpc$. }

\subsection{
Constraining $M$ and $R$ with $\rinfty$ and $A$: the
%Evaluation of  NS  mass and  radius by
% $\rinfty$ and A: application
case of Aql X-1}

\begin{figure}
\includegraphics[width=9cm]{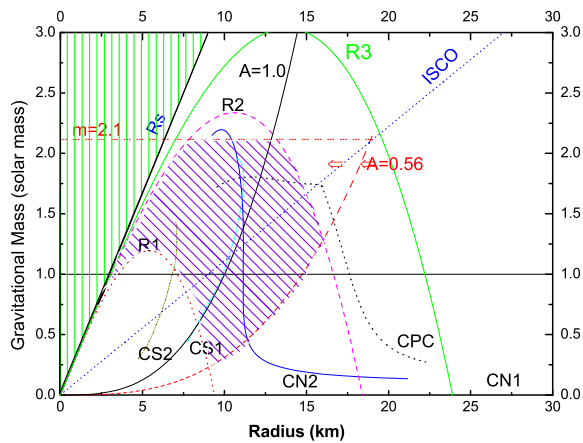}
%
%\vskip 6.5cm \hskip   -0.40cm \epsfxsize=7.cm
%\special{psfile=f2.eps hscale=90 vscale=90 angle= 0} \hskip -1.5cm
%
 \caption[fig2] {\bf  Similar to Figure \ref{iz}.
 The mass and  radius of   NS in Aql X-1
are constrained in the shaded area. R1 (R2) represents $\rinfty=7.5
(23.9)$ (km), where  the lower and upper limits of \rinfty on
account of the uncertainty by the  source distance are 7.5 km ((13.4
- 4)* 4/5) and 23.9 km ((13.4+5)*6.5/5), respectively. The two
concave-up parabolas stand for $A=0.56$ and $A=1.0$, respectively
(as marked).  The horizontal line $m=2.1$ represents M=2.1\ms. }
\label{ia}
\end{figure}

If the apparent radius $\rinfty$  and
the parameter $A$ are known, both $M$ and $R$ can be calculated by the
following two formulae, derived from Eqs.(\ref{rinfty}) and
(\ref{mr}),
  \be
\rsix=\rinfsix/\sqrt{1+0.15(A/0.7)^{2}R^{2}_{\infty6}}\;,
\label{ria}\ee \be m=0.49(A/0.7)^{2}
\left(\rinfsix/\sqrt{1+0.15(A/0.7)^{2}R^{2}_{\infty6}}\right)^{3}\;.
\label{mia}\ee {\bf The quiescent spectrum of the transient type-I
X-ray bursting NS Aql X-1
 is measured with Chandra/ACIS-S,  and the best-fit value of
the  apparent radius
\rinfty=13.4\ud{5}{4} $(d/5 kpc)$ km is inferred by  assuming a
pure hydrogen atmosphere with the temperatures ranging from 145--168 eV,
plus a power-law component.
The distance from the observer to the source is  5 kpc as a fiducial
value because the current uncertainties allow
for a distance from 4  kpc to  6.5 kpc (Rutledge et al. 2001). Therefore,
  including the uncertainty on   source distance the lower and upper
limits on \rinfty are 7.5 km ((13.4 - 4)* 4/5)  and 23.9 km
((13.4+5)*6.5/5), respectively.   Moreover, the single kHz QPO
frequency 1040 Hz has been detected in Aql X-1 by RXTE (e.g. van der
Klis 2000, 2006), so that according to Eqs.
%the parameter A and mass m can be
%constrained  by the QPO frequency  from  Eq.
(\ref{a}) and (\ref{mass2}), one can get the constraints $A\ge0.56$
and $m\le2.1$.  In Figure \ref{ia}, the ranges of $M$ and $R$ of  NS
in Aql X-1 are confined in the shadowed area where the meanings of
the boundaries are indicated in the figure. We find that the
shadowed area has a loose constraint on the NS EOSs, and it covers
 the representative chosen  EOSs of the strange matter, the
 normal  neutrons and the pion condensate in the star core.
Moreover, it is worth noticing that the $A=1.0$ parabola divides the
$M-R$ diagram into two parts (Figure. \ref{ia}), and the strange
star EOSs are relevant only when $A>1.0$

}
%do the strange star EOSs
% only above which the star EOS may be involved in the exotic matter.

%670 Hz, 930 Hz

\subsection{
Constraining $M$ and $R$ with $A$ and $z$: the case
%Evaluation of  NS  mass and  radius by A and z:
% application case
of EXO 0748-676}

If the gravitational redshift $z$ and the parameter $A$ are known,
both $M$ and $R$ can be calculated by the following two formulae,
derived from Eqs.(\ref{redshift}) and (\ref{mr}),
\be
R_{6} = 1.43 f^{1/2}(z)({A/0.7})^{-1}\;,
\label{radius} \ee \be m = 1.43 f^{3/2}(z) ({A/0.7})^{-1}\;.
\label{mass}
\ee
In case that only a single kHz QPO frequency is
detected, we can place a lower limit on the parameter $A$ from
Eq.(\ref{a}). Following Eqs.(\ref{radius}) and (\ref{mass}), $M$ and
$R$ are constrained by the following inequalities
%, from
%Eq.(\ref{radius}) and Eq.(\ref{mass}), %as well as Eq.(\ref{a}),
\be
 R_{6} \le
  (1850 ({\rm Hz})/\nk)f^{1/2}(z)   \,, \label{radius3}
\ee
\be
m \le   (1850 ({\rm Hz})/\nk)f^{3/2}(z)  \,. \label{mass3}
\ee

\begin{figure}
\includegraphics[width=9cm]{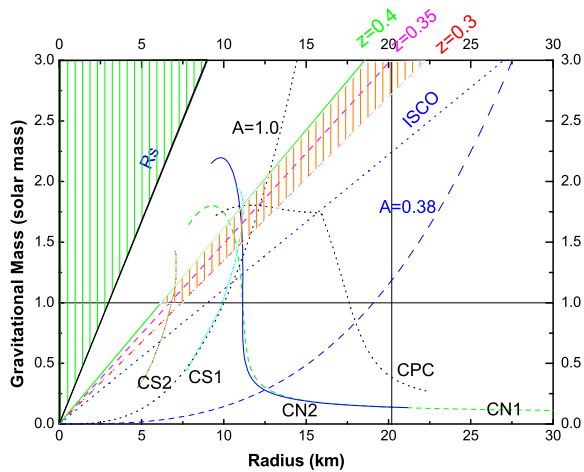}
%
%\vskip 6.5cm \hskip   -0.40cm \epsfxsize=7.cm
%\special{psfile=f3.eps hscale=90 vscale=90 angle= 0} \hskip -1.5cm
%
\caption[fig1] {Similar to
 Figures \ref{iz} and \ref{ia}. The mass and   radius
%estimations of star
of the NS in EXO 0748-676 are constrained by the line $z=0.35$ and
two  horizontal  lines representing   one (three) solar mass(es).
 Cottam et al. (2002) declaimed the gravitational redshift $z=0.35$ with
 small error, but this samll error is not given quantitatively. As a
 possible comparison, we plot the lines with the redshift
  $z=0.3$ and z=0.4.  }
\label{az}
\end{figure}

% As for the application to the particular source,
{\bf
A gravitational redshift $z=0.35$ was
 detected  in EXO 0748-676 by Cottam et al. (2002),
 who identify the most significant features with the Fe XXVI
  and XXV n=2-3 and O VIII
n=1-2 transitions, all with a redshift of $z=0.35$, identical within
small uncertainties for the respective transitions, however this
small error has not been given  quantitatively by the authors. In
order to present the possible influence of the redshift variation on
the $M-R$ region, we plot $z=0.3$ and $z=0.4$ lines in Figure
\ref{az}. }
%where a
A single kHz QPO frequency 695 Hz is detected by RXTE for the same
source (e.g., Homan \& van der Klis 2000; van der Klis 2000). Thus
$A\ge0.38$ and $m\le3.2$ are inferred from Eq.(\ref{a}) and
Eq.(\ref{mass2}). As shown in Figure \ref{az}, the $A\ge 0.38$
condition gives a very loose constraint,  {\bf which has little help
to constrain the NS mass into a domain of less than 3 $\ms$ because
the detected QPO frequency  of this source is abnormally lower than
the typical values 1000 Hz (see, van der Klis 2000, 2006).  If the
future detections of this source
 present the higher frequency, then the promising constrain condition
would be improved.}
Moreover, since M=$0.97\pm0.24\ms$ was measured as
a lower mass for 2A 1822 - 371 (Jonker et al 2003),
we set
%a conservative
%from the currently accepted astrophysical arguments, the
%conservative lower and upper limit of NS mass can be  set  as
1.0 $\ms $ and 3.0 $\ms$ as the conservative lower and upper mass
limits for NSs, respectively.
%because M=$0.97\pm0.24\ms$ as
%a mass lower limit was  measured  in   2A 1822 - 371 (Jonker et al
%2003),  therefore the
The $M-R$ constraint for EXO 0748-676 is then limited to the segment
of  the straight dashed line $z=0.35$  between the two horizontal
lines $m=1.0$ and $m=3.0$, where the EOSs for the strange matter,
normal neutron and pion condensate in the star core  are all
permitted.
% Moreover,  as for the detected high
%gravitational redshift z=0.35  in  EXO 0748-676 (Cottam et al.
%2002), the possible explanation is that the position to exhibit
%the X-ray spectrum locates inside ISCO  or its star radius
%satisfies  $R < \ris$  because   ISCO line corresponds to  the
%gravitational redshift z=0.23.

\section{Summary and discussion}

In this paper, we discuss the possibility of using any two of the
three $M-R$ relationships to constrain NS mass and radius. The three
$M-R$ relationships include
%have studied the evaluations of NS mass and
%radius by any two of the three possibly known M-R  relations
%derived from
the ``apparent radius" ($\rinfty$), the gravitational redshift ($z$)
and the averaged mass density ($A^2$).
%of star, together with  the demonstrations of
We applied the method to three sample sources, 1E1207.4-5209, Aql X-1
and EXO 0748-676. In principle, if all the above three $M-R$ relations
are well measured in one source, one can pose tight constraints on the
ranges of $M$ and $R$, so that
%relations  exist in one source,  the ranges of the independent
%values of M and R can be properly  implied in M-R diagram and the
the potential NS EOSs would be inferred. Unfortunately, this is not
achieved so far.
%at the present
%stage, there does not exist a source possessing  these  three
%measures of M-R relations.
Furthermore, some $M-R$ relations are incomplete or have large
errors, so that additional
%if the existed measures
%of M-R relations are incomplete or have big errors, we may
%construct the constrain conditions for the M-R relations  from the
astrophysical arguments are needed, and one can derive
%such as
some inferred lower or upper
limits of the $M-R$ relations
%(or M) described in
(e.g. Eqs.[\ref{a}] and [\ref{mass2}]).
In binary systems, if the NS mass is known,
% In addition to the known NS  mass  measured in the binary system,
any one of the three $M-R$ relation measurements would be sufficient
to derive its radius.

For the compact star in 1E1207.4-5209, the estimates of its mass and
radius come from the possible lower limit of the calculated apparent
radius of about 4.6 km (Bignami et al. 2004). An extraordinary low
mass is inferred, which is even less than the measured  lower limit
of NS mass $0.97\pm0.24 \ms$ in  2A 1822 - 371 (Jonker et al. 2003).
In such a case, the existence of exotic matter inside the star is
hinted under the assumptions of blackbody spectrum and $2kpc$
distance.

The stellar radius can be also estimated by the motion-induced
Doppler-broadening of absorption lines, when the emission region moves
towards and away from the observer as the star rotates.
According to this, Villarreal and Strohmayer (2004) have studied
the source EXO 0748-676 with the measured gravitational redshift
$z=0.35$ (Cottam et al. 2002),
 and obtained a radius of about 11.5 km and
a stellar mass of about 1.75 \ms.
A remark is that the emission latitude
 and the orientation of the rotation axis both affect the result,
so that the method is model-dependent.
In addition,  from the recent further study,
Chang et al. (2005) show
 that the intrinsically broad line profile prohibits
any meaningful constraint on the NS radius if the 45 Hz burst
oscillation seen by Villarreal \& Strohmayer (2004) is the spin
frequency.

 It is worth to mention the measured  apparent radius of isolated
NS in \rxj  {\;} because the distance of this source to earth is
well determined to be about 117 pc (see, e.g. Walter \& Lattimer
2002), with the perfect black-body spectrum of this source (Drake et
al. 2002; Burwitz et al. 2003, 2001)  the conservative lower limit
of apparent radius has been implied to be 16.5 km (d/117 pc)
(Truemper et al. 2004). This corresponds to the   ''true`` radius of
14\,km for a 1.4\,M$_{\odot}$ neutron star, indicating a stiff
equation of state at high densities, which excludes the quark stars
or even the NSs with the quark cores (Truemper et al. 2004).
Moreover, without knowledge of the spectrum line in \rxj {\; } until
now (see e.g. Drake et al. 2002), we cannot constrain its M and R
independently (Truemper et al. 2005). {\bf Once again,  it is noted
that the above conclusion comes from the black-body spectrum
assumption, and
 the apparent radius may confront a modification by more than a factor of 2
 if using the atmosphere model of photon  spectrum (Haensel 2001). }

%In conclusion,
The next generation X-ray observatories with enhanced
%on the spectra and timing
spectral and timing capabilities would greatly improve the
``measurements" of the three $M-R$ relations discussed in this paper.
This would strengthen
%deriving the star  M-R  relations, and it would be perspective to
%raise
our confidence in evaluating the super-dense nuclear
compositions of compact stars.
Theoretically, if both $M$ and $R$
of a compact star are accurately measured, its EOS and nuclear matter
compositions would be explicitly revealed. This in turn sheds light on
the unknown
%even hints the
astrophysical processes during the supernova explosion of its
progenitor star (e.g. Podsiadlowski et al. 2005). On the other hand,
the regime of non-Newtonian strong gravitational fields would be
also revealed with the determination of the NS mass and radius,
where the  direct tests of Einstein's theory of general relativity
will be possible in the near future (e.g. van der Klis 2000, 2006).

 \vskip 0.5cm

%\begin{acknowledgements}
\section{acknowledgements}

  We are grateful for M.C. Miller for providing the EOS data files.
  Thanks  are also due to the helpful discussions with V. Burwitz,
 J. Truemper, J.M. Lattimer.  We  especially thank J. Truemper
 for many suggestions and advices, and for initialing discussions on
  M-R relation of \rxj.  Thanks are also duo to the helpful discussions
  with K.S. Cheng, Z.G. Dai, D.M. Wei, T. Lu, Q.H. Peng, Z.R. Wang,
  J.L. Han, G.J. Qiao, and X.J. Wu. This research has been
  supported by the  innovative project of CAS of   China.
 We are very grateful for the critic comments from the anonymous referee
that greatly improve the  quality of the paper.
%on the English writing of  the manuscript.

Note added: after acceptance of the paper, we paid attention to the
work by F. Ozel on the mass and radius of NS in EXO 0748-676 (Nature
2006, 441, 1115)

%\end{acknowledgements}

\end{document}